\documentclass[pra,aps,amsmath,amssymb,showkeys]{revtex4}
\newcommand{\bro}{{\hat{\rho}}}
\newcommand{\vbro}{{\hat{\varrho}}}
\newcommand{\bsg}{{\hat{\sigma}}}
\newcommand{\bomg}{{\hat{\omega}}}

\newcommand{\taug}{{\hat{\tau}}}
\newcommand{\Tr}{{\mathrm{Tr}}}
\newcommand{\itt}{{\mathtt{i}}}
\newcommand{\xdif}{{\mathrm{d}}}
\newcommand{\amh}{{\hat{A}}}
\newcommand{\bmh}{{\hat{B}}}
\newcommand{\mmh}{{\hat{M}}}
\newcommand{\nnh}{{\hat{N}}}
\newcommand{\skx}{{\hat{k}}}
\newcommand{\sqx}{{\hat{q}}}

\newcommand{\sax}{{\hat{x}}}
\newcommand{\lap}{{\hat{\Lambda}}}
\newcommand{\laq}{{\hat{\Pi}}}
\newcommand{\cla}{{\mathcal{A}}}
\newcommand{\clb}{{\mathcal{B}}}
\newcommand{\clk}{{\mathcal{K}}}
\newcommand{\clm}{{\mathcal{M}}}
\newcommand{\cln}{{\mathcal{N}}}

\newcommand{\clx}{{\mathcal{X}}}
\newcommand{\pq}{{\mathtt{p}}}

\begin{document}
\clearpage
\preprint{}

\title{Entropic uncertainty relations for successive measurements in the presence of a minimal length}

\author{Alexey E. Rastegin}

\affiliation{Department of Theoretical Physics, Irkutsk State University,
Gagarin Bv. 20, Irkutsk 664003, Russia}

\begin{abstract}
We address the generalized uncertainty principle in
scenarios of successive measurements. Uncertainties are
characterized by means of generalized entropies of both the
R\'{e}nyi and Tsallis types. Here, specific features of
measurements of observables with continuous spectra should be
taken into account. First, we formulated uncertainty relations in
terms of Shannon entropies. Since such relations involve a
state-dependent correction term, they generally differ from
preparation uncertainty relations. This difference is revealed
when position is measured by the first. In contrast,
state-independent uncertainty relations in terms of R\'{e}nyi and
Tsallis entropies are obtained with the same lower bounds as in
the preparation scenario. These bounds are explicitly dependent on
the acceptance function of apparatuses in momentum measurements.
Entropic uncertainty relations with binning are discussed as well.
\end{abstract}


\maketitle

\pagenumbering{arabic}
\setcounter{page}{1}

\section{Introduction}\label{sec1}

The Heisenberg uncertainty principle \cite{heisenberg} is now
avowed as a fundamental scientific concept. Heisenberg
examined his thought experiment rather qualitatively. An explicit
formal derivation appeared in \cite{kennard}. This approach
was later extended to arbitrary pairs of observables \cite{robert}.
These traditional formulations are treated as preparation
uncertainty relations \cite{rozp17}, since repeated trials with
the same quantum state are assumed here. This simplest scenario
differs from the situations typical in quantum information science.
Since uncertainty relations are now examined not only
conceptually, researchers often formulated them in
information-theoretic terms. As was shown in \cite{ckw14},
wave-particle duality can be interpreted on the basis of entropic
uncertainty relations. Basic developments within the entropic
approach to quantum uncertainty are reviewed in
\cite{ww10,brud11,cbtw17}. Interest in this approach has been
stimulated by advances in using quantum systems as an
informational resource
\cite{renner10,renner11,nbw12,furrer14,lfei18}. Among more
realistic cases, scenarios with successive measurements have been
addressed in the literature
\cite{mdsrin03,paban13,bfs2014,zzhang14,bs2016}. Researchers are
currently able to manipulate individual quantum systems
\cite{wineland13,haroche13}. In quantum information processing,
our subsequent manipulations usually deal with an output of a
latter stage. In effect, Heisenberg's thought experiment with
microscope should rather be interpreted as related to
uncertainties in successive measurements \cite{blw13}. Uncertainty
relations in the scenarios of successive measurements have
received less attention than they deserve \cite{paban13}. The
authors of \cite{paban13} also compared their findings with
noise-disturbance relations given in \cite{ozawa04}. Studies of
scenarios with successive measurements allow us to understand
whether preparation uncertainty relations are applicable to one or
another question.

In principle, the Heisenberg uncertainty principle does not impose
a restriction separately on spreads of position and momentum. It
merely reveals that continuous trajectories are unspeakable in
standard quantum mechanics, although such principles remain valid within
Bohmian mechanics \cite{gisin18}. The generalized uncertainty
principle is aimed to involve the existence of a minimal
observable length. The latter is naturally connected with efforts
to describe quantum gravity \cite{hossen13}. Some advances in
merging quantum mechanics and general relativity are summarized in
\cite{rovelli04}. It is believed that quantum gravitational
effects begin to be apparent at the scale corresponding to the
Planck length
$\ell_{P}=\sqrt{G\hbar/c^{3}}\approx1.616\times10^{-35}$ m. Below
this scale, the very structure of space-time is an open problem
\cite{amati89}. In addition, Heisenberg's principle is assumed to
be converted into the generalized uncertainty principle (GUP)
\cite{scard99,bombi08,td2015}. There exist proposals to test
observable effects of the minimal length, including astronomical
observations \cite{ellis98,piran07} and experimental schemes
feasible within current technology \cite{brukner12,ffm13}. The GUP
case connects to many aspects that are currently the subject of
active researches \cite{tawfik13,deys13,diab14,faizalm15,mfza16}.
The generalized uncertainty principle declares a non-zero lower
bound on position spread. To reach such a model, the canonical
commutation relation should be modified. Deformed forms of the
commutation relation were recently studied from several
viewpoints. On the other hand, the connections of the GUP with the
real world represent an open question. In the context of
non-relativistic quantum mechanics, the corresponding formalism
was proposed in \cite{kempf95}. Another approach to representation
of the used observables was suggested in \cite{pedram12}. This way
is very convenient in extending entropic uncertainty relations to
the GUP case \cite{pedram16}.

In this paper, we aim to consider entropic uncertainty relations
for successive measurements in the presence of a minimal
observable length. Of course, our presentation is essentially
based on mathematical relations given by Beckner \cite{beck} and
by Bia{\l}ynicki-Birula and Mycielski \cite{birula1}. This
direction was initially inspired by Hirschman \cite{hirs}. For
observables with finite spectra, basic developments appeared due
to \cite{deutsch,kraus,maass}. We will largely use the results
reported in \cite{cancon16,gegen17}. The work in \cite{cancon16} is
devoted to formulating entropic uncertainty relations for
successive measurements of canonically conjugate observables. The
case of position and momentum was addressed therein as a
particular example of the scheme developed in \cite{pvb90,gvb95}.
Entropic uncertainty relations in the presence of a minimal length
were examined in \cite{gegen17}, and mainly focused on those
points that were not considered in this context previously.
Combining these two aspects finally led to the generalized
uncertainty principle in scenarios of successive measurements. This
paper is organized as follows. In Section \ref{sec2}, we review
preliminary material, including properties of used
information-theoretic measures. In Section \ref{sec3}, we briefly
discuss successive quantum measurements in general. The main results
of this paper are presented in Section \ref{sec4}. Both of the
typical scenarios of successive measurements will be examined. In
particular, we will see how formulating lower entropic bounds
depends on the actual order in which measurements of position and
momentum have been performed. In Section \ref{sec5}, we conclude
the paper with a summary of the obtained results.

\section{Preliminaries}\label{sec2}

In this section, we review the required material and fix the
notation. To characterize measurement uncertainties, we use
entropies of the R\'{e}nyi and Tsallis types. Let us begin with
the case of probability distributions with a discrete label. For
the given probability distribution $\pq=\{p_{i}\}$, its R\'{e}nyi
entropy of order $\alpha$ is defined as \cite{renyi61}
\begin{equation}
R_{\alpha}(\pq):=\frac{1}{1-\alpha}\>\ln\left(\sum\nolimits_{i} p_{i}^{\alpha}\right)
, \label{rpdf}
\end{equation}
where $0<\alpha\neq1$. For $0<\alpha<1$, the R\'{e}nyi
$\alpha$-entropy is a concave function of the probability
distribution. For $\alpha>1$, it is neither purely convex nor
purely concave \cite{jizba}. In the limit $\alpha\to1$, the
formula (\ref{rpdf}) gives the standard Shannon entropy
\begin{equation}
H_{1}(\pq)=-\sum\nolimits_{i}p_{i}\ln{p}_{i}
\, . \label{shan}
\end{equation}
For the given probability distribution $\pq=\{p_{i}\}$ and
$0<\alpha\neq1$, the Tsallis $\alpha$-entropy is defined as
\cite{tsallis}
\begin{equation}
H_{\alpha}(\pq):=\frac{1}{1-\alpha}\,
\left(\sum\nolimits_{i} p_{i}^{\alpha}-1\right)
=-\sum\nolimits_{i}p_{i}^{\alpha}\ln_{\alpha}(p_{i})
 . \label{tsent}
\end{equation}

Here, we use the $\alpha$-logarithm expressed as
$\ln_{\alpha}(y):=\bigl(y^{1-\alpha}-1\bigr)/(1-\alpha)$ for
positive variable $y$ and $0<\alpha\neq1$. When $\alpha\to1$, the
$\alpha$-logarithm reduces to the usual one. Then, the
$\alpha$-entropy (\ref{tsent}) also leads to the Shannon entropy
(\ref{shan}). An axiomatic approach to generalized
information-theoretic quantities is reviewed in \cite{imrec08}. In
more detail, properties and applications of generalized entropies
in physics are discussed in \cite{bengtsson}. In the present
paper, we will deal only with entropies of probability
distributions. Quantum entropies of very general family were
thoroughly examined in \cite{hbb15,bzhpl16}. Quantum R\'{e}nyi and
Tsallis entropies are both particular representatives of this
family.

Let $w(x)$ be a probability density function defined for all real
$x$. Then, the differential Shannon entropy is introduced as
\begin{equation}
H_{1}(w):=-\int_{-\infty}^{+\infty} w(x)\,\ln{w}(x)\,\xdif{x}
\, . \label{denwy}
\end{equation}

Similarly, we determine entropies for other continuous variables
of interest. For $0<\alpha\neq1$, the differential R\'{e}nyi
$\alpha$-entropy is defined as
\begin{equation}
R_{\alpha}(w):=\frac{1}{1-\alpha}\>
\ln\!\left(\int_{-\infty}^{+\infty} w(x)^{\alpha}\,\xdif{x}\right)
 . \label{reddf}
\end{equation}

In contrast to entropies of a discrete probability distribution,
differential entropies are not positive definite in general. To
quantify an amount of uncertainty, we often tend to deal with
positive entropic functions. One possible approach is such
that the continuous axis of interest is divided into a set of
non-intersecting bins. Preparation uncertainty relations with
binning were derived in terms of the Shannon \cite{IBB84} and
R\'{e}nyi entropies \cite{IBB06}. To reach a good exposition, the
size of these bins should be sufficiently small in comparison with
a scale of considerable changes of $w(x)$. Keeping an obtained
discrete distribution, we further calculate entropies of the forms
(\ref{rpdf}) and (\ref{tsent}).

The generalized uncertainty principle declares the deformed
commutation relation for the position and momentum operators
\cite{kempf95}. For convenience, we will use the wavenumber
operator $\skx$ instead of the momentum operator $\hbar\skx$. It
is helpful to rewrite this relation as
\begin{equation}
\bigl[\sax,\skx\bigr]=\itt\bigl(1+\beta\skx^{2}\bigr)
\, . \label{gcomr}
\end{equation}

Here, the positive parameter $\beta$ is assumed to be rescaled by
factor $\hbar^{2}$ from its usual sense. With the limit $\beta\to0$,
the formula (\ref{gcomr}) gives the standard commutation relation
of ordinary quantum mechanics. Due to the Robertson formulation
\cite{robert}, the standard deviations in the pre-measurement
state $\bro$ satisfy
\begin{equation}
\Delta\amh\,\Delta\bmh\geq
\Bigl|\frac{1}{2}\>\bigl\langle[\amh,\bmh]\bigr\rangle_{\bro}\Bigr|
\, . \label{robfor}
\end{equation}

By $\langle\amh\rangle_{\bro}=\Tr(\amh\,\bro)$, we mean the
quantum-mechanical expectation value. Combining (\ref{gcomr}) with
(\ref{robfor}) then gives
\begin{equation}
\Delta\sax\,\Delta\skx\geq
\frac{1}{2}
\,\bigl(1+\beta\langle\skx^{2}\rangle_{\bro}\bigr)
\geq\frac{1}{2}
\,\bigl(1+\beta(\Delta\skx)^{2}\bigr)
\, . \label{robr}
\end{equation}

The principal parameter $\beta$ is positive and independent of
$\Delta\sax$ and $\Delta\skx$ \cite{kempf95}. It directly follows
from (\ref{robr}) that $\Delta\sax$ is not less than the square
root of $\beta$. As was shown in \cite{pedram12}, the auxiliary
wavenumber operator $\sqx$ allows us to mediate between
(\ref{gcomr}) and the standard commutation relation. Let $\sax$
and $\sqx$ be self-adjoint operators satisfying
$[\sax,\sqx]=\itt$. In the $q$-space, the action of $\sqx$
results in multiplying a wave function $\varphi(q)$ by $q$,
whereas $\sax\varphi(q)=\itt\,\xdif\varphi/\xdif{q}$. Then, the
wavenumber $\skx$ can be represented as \cite{pedram12}
\begin{equation}
\skx=\frac{1}{\sqrt{\beta}}\>\tan\bigl(\sqrt{\beta}\sqx\bigr)
\, . \label{momr}
\end{equation}

The auxiliary wavenumber obeys the standard commutation relation
but ranges between $\pm\,q_{0}(\beta)=\pm\,\pi/(2\sqrt{\beta}\,)$.
The function $q\mapsto{k}=\tan(\sqrt{\beta}q)\big/\sqrt{\beta}$
gives a one-to-one correspondence between
$q\in(-\,q_{0};+\,q_{0})$ and $k\in(-\,\infty;+\,\infty)$. So, the
eigenvalues of $\skx$ fully cover the real axis. Further details
of the above representation are examined in \cite{pedram12}.

For any pure state, we will deal with three wave functions
$\phi(k)$, $\varphi(q)$, and $\psi(x)$. The formalism of
\cite{pedram12} is convenient in the sense that it explicitly
describes the space of acceptable wave packets. In the $q$-space,
these states should have wave functions that vanish for
$|q|>q_{0}(\beta)$. Here, the auxiliary wave function $\varphi(q)$
is a useful tool related to $\psi(x)$ via the Fourier transform.
In the $q$-space, the eigenfunctions of $\sax$ appear as
$\exp(-\itt{q}x)\big/\sqrt{2\pi}$. Thus, any wave function in the
coordinate space is expressed as
\begin{equation}
\psi(x)=\frac{1}{\sqrt{2\pi}}\int_{-q_{0}}^{+q_{0}}
\exp(+\itt{q}x)\,\varphi(q)\,\xdif{q}
\, . \label{psxdf}
\end{equation}

Wave functions in the $q$- and $x$-spaces are connected by
the Fourier transform \cite{pedram12},
\begin{equation}
\varphi(q)=\frac{1}{\sqrt{2\pi}}\int_{-\infty}^{+\infty}
\exp(-\itt{q}x)\,\psi(x)\,\xdif{x}
\, . \label{phpdf}
\end{equation}

The distinction from ordinary quantum mechanics is that wave
functions in the $q$-space should be formally treated as $0$ for
all $|q|>q_{0}(\beta)$.

Using the above connection, the author of \cite{pedram16} affirmed
the following. The uncertainty relation given in
\cite{beck,birula1} is still valid in the GUP case. However, wave
functions in the $q$-space are actually auxiliary. In the GUP
case, the physically legitimate wavenumber and momentum involved
in the relation (\ref{gcomr}) are described by wavefunctions in
the $k$-space. A real distribution of physical wavenumber values
is determined with respect to $\phi(k)$ instead of $\varphi(q)$.
Let us examine the probability that momentum lies between two
prescribed values. In view of the bijection between the intervals
$(k_{1};k_{2})$ and $(q_{1};q_{2})$, this probability is expressed
as
\begin{equation}
\int_{k_{1}}^{k_{2}} |\phi(k)|^{2}\,\xdif{k}=
\int_{q_{1}}^{q_{2}} |\varphi(q)|^{2}\,\xdif{q}
\, , \label{twopr}
\end{equation}
so that $|\phi(k)|^{2}\,\xdif{k}=|\varphi(q)|^{2}\,\xdif{q}$.
Hence, two probability density functions $u(k)$ and $v(q)$ are
connected as $u(k)\,\xdif{k}=v(q)\,\xdif{q}$, in another form
\begin{equation}
u(k)=\frac{v(q)}{1+\beta{k}^{2}}
\, . \label{upvq}
\end{equation}

For pure states, when $u(k)=|\phi(k)|^{2}$ and
$v(q)=|\varphi(q)|^{2}$, the formula (\ref{upvq}) is obvious. It
can be extended to mixed states due to the spectral
decomposition. However, one is actually unable to obtain the
probability density functions $u(k)$ and $w(x)$ immediately.

In reality, any measurement apparatus is inevitably of a finite
size. Devices with a finite extension need a finite amount of
energy. Hence, one cannot ask for some state in which the
measurement of an observable gives exactly one particular value of
position. In more detail, measurements of coordinates of a
microparticle are considered by Blokhintsev (\cite{blokh73}, Chapter II). The
generalized uncertainty principle imposes another limitation for
position measurements. Although eigenstates of position and
momentum are often considered explicitly, they are rather
convenient tools of mathematical technique. The corresponding kets
are not elements of the Hilbert space, but can be treated in the
context of rigged Hilbert spaces \cite{mbg02}. Instead, we aim to
use narrow distributions of a finite but small width. Measuring or
preparing some state with the particular value $\xi$ of position,
one has to be affected by a neighborhood of $\xi$. Therefore, we
treat each concrete result only as an estimation compatible
with the GUP.

Thus, we cannot directly obtain probability density functions of
the form $u(k)$ and $w(x)$. Here, a finiteness of detector
resolution should be addressed \cite{paban13,cancon16}. Measuring
or preparing a state with the particular value $\xi$ of position,
one is affected by some vicinity of $\xi$. In this way, we refer
to generalized quantum measurements. Let the eigenkets $|x\rangle$
be normalized through Dirac's delta function. As was already
mentioned, such kets cannot be treated as physical states even
within ordinary quantum mechanics. In a finite-resolution
measurement of position, the set
$\clx=\bigl\{|x\rangle\langle{x}|\bigr\}$ is replaced with some
set $\cln$ of operators of the form
\begin{equation}
\nnh(\xi):=\int_{-\infty}^{+\infty}\xdif{x}\,g(\xi-x)\,|x\rangle\langle{x}|
\, . \label{nnxi}
\end{equation}

An acceptance function $\xi\mapsto{g}(\xi)$ satisfies the
condition $\int_{-\infty}^{+\infty}|g(\xi)|^{2}\,\xdif\xi=1$.
Then operators of the form (\ref{nnxi}) lead to a generalized
resolution of the identity,
\begin{equation}
\int_{-\infty}^{+\infty}\xdif\xi\,\nnh(\xi)^{\dagger}\nnh(\xi)=1
\, , \label{csrl}
\end{equation}
where the right-hand side is treated as the identity operator.
For the pre-measurement state $\bro$, the measurement leads to the
probability density function
\begin{equation}
W_{\bro}(\xi)=\Tr\bigl(\nnh(\xi)^{\dagger}\nnh(\xi)\bro\bigr)
=\int_{-\infty}^{+\infty}|g(\xi-x)|^{2}\,w_{\bro}(x)\,\xdif{x}
\, . \label{bigw}
\end{equation}

This should be used instead of
$w_{\bro}(x)=\langle{x}|\bro|x\rangle$. When the acceptance
function is sufficiently narrow, we will obtain a good
``footprint'' of $w_{\bro}(x)$. Let $\zeta\mapsto{f}(\zeta)$ be
another acceptance function that also obeys the normalization
condition. A finite-resolution measurement of the legitimate
wavenumber is described by some set $\clm$ of operators
\begin{equation}
\mmh(\zeta):=\int_{-\infty}^{+\infty}\xdif{k}\,f(\zeta-k)\,|k\rangle\langle{k}|
\, . \label{mmzt}
\end{equation}

Here, the initial resolution
$\clk=\bigl\{|k\rangle\langle{k}|\bigr\}$ is replaced with
$\clm=\bigl\{\mmh(\zeta)\bigr\}$. Instead of
$u_{\bro}(k)=\langle{k}|\bro|k\rangle$, we actually deal with the
probability density function
\begin{equation}
U_{\bro}(\zeta)=\Tr\bigl(\mmh(\zeta)^{\dagger}\mmh(\zeta)\bro\bigr)
=\int_{-\infty}^{+\infty}|f(\zeta-k)|^{2}\,u_{\bro}(k)\,\xdif{k}
\, , \label{bigu}
\end{equation}

For good acceptance functions, a distortion of statistics will be
small. The Gaussian distribution is a typical form of such
functions \cite{paban13}. We will assume that a behavior of
acceptance functions is qualitatively similar.

\section{On successive measurements of observables in general}\label{sec3}

In this section, we generally formulate the question with respect
to two successive measurements of observables with continuous
spectra. It is more sophisticated than an intuitive obvious
treatment of successive measurements on a finite-dimensional
system. The latter allows us to deal with projective measurements,
since all observables have a discrete spectrum. Such an approach is
not meaningful for the case of position and momentum. On the other
hand, the finite-dimensional case is important for understanding
basic formulations related to continuous observables. To motivate
our approach, we briefly review entropic uncertainty relations
for successive projective measurements. Further, we will present a
suitable reformulation for the case of position and momentum.
Together with the entropic formulation, other approaches to
express uncertainties in quantum measurements are of interest. In
particular, modern investigations are based on the sum of
variances \cite{huang12,mpati14}, majorization relations
\cite{prz13,fgg13,rpz14,lrud15,arkz16}, and the method of effective
anticommutators \cite{ktw14}. The authors of \cite{lbp16}
discussed some surprising results that may occur in application of
entropic measures to quantify uncertainties in quantum
measurements. These questions are beyond the scope of our
consideration.

Scenarios with successive measurements are of interest for several
reasons. The concept of wave function reduction assumes that we
perform at least two successive measurements on a system (see for
example Section 5.5 of \cite{bdqm02}). By $\lap_{a}\in\cla$, we denote
a projector onto the $a$-th eigenspace of finite-dimensional
observable $\amh$. For the pre-measurement state $\bro$, the
probability of outcome $a$ is written as $\Tr(\lap_{a}\bro)$.
Such probabilities form a discrete distribution, from which
we calculate quantities of interest. By $R_{\alpha}(\cla;\bro)$
and $H_{\alpha}(\cla;\bro)$, we further mean the entropies
(\ref{rpdf}) and (\ref{tsent}) calculated with the probabilities
$\Tr(\lap_{a}\bro)$. After the measurement of $\amh$, we measure
another observable $\bmh$. It is actually described by the set
$\clb=\{\laq_{b}\}$. Note that subsequent measurements are assumed
to be performed with a new ensemble of states. The latter differs
from traditional uncertainty relations in the preparation
scenario. Scenarios with successive measurement are fixed by the
used form of post-first-measurement states \cite{bfs2014}.

In the first scenario, the second measurement is performed on the
state immediately following the first measurement with completely
erased information. Here, the pre-measurement state of the second
measurement is expressed as \cite{mdsrin03}
\begin{equation}
\Upsilon_{\cla}(\bro)=\sum\nolimits_{a}\lap_{a}\bro\lap_{a}
\ . \label{fsmpm}
\end{equation}

To characterize the amount of uncertainty in two successive
measurements, we will use quantities of the form
\begin{equation}
R_{\alpha}(\cla;\bro)+R_{\gamma}\bigl(\clb;\Upsilon_{\cla}(\bro)\bigr)
\, , \label{rescp}
\end{equation}
and similarly with the corresponding Tsallis entropies. In the
second scenario of successive measurements, we assume that the
result of the first measurement is maintained. A focus on actual
measurement outcomes is typical for the so-called selective
measurements. For example, incoherent selective measurements are
used in the formulation of monotonicity of coherence measures
\cite{bcp14}. Coherence quantifiers can be defined with entropic
functions of the Tsallis \cite{rast16a} and R\'{e}nyi
types \cite{zhc17}. In effect, the second measurement will be
performed on the post-first-measurement state selected with
respect to the actual outcome \cite{bfs2014,zzhang14}. Due
to the L\"{u}ders reduction rule \cite{luders51}, this state
is written as
\begin{equation}
\taug_{a}=\bigl(\Tr(\lap_{a}\bro)\bigr)^{-1}\,\lap_{a}\bro\lap_{a}
\, , \label{ludr}
\end{equation}
whenever $\Tr(\lap_{a}\bro)\neq0$. Measuring the observable
$\bmh$ in each $\taug_{a}$, we obtain the corresponding entropy
$R_{\gamma}(\clb;\taug_{a})$. Averaging over all $a$, we introduce
the quantity
\begin{equation}
\sum\nolimits_{a} \Tr(\lap_{a}\bro)\, R_{\gamma}(\clb;\taug_{a})=
\sum\nolimits_{a} \Tr(\lap_{a}\bro)\, R_{\alpha}(\cla;\taug_{a})
+\sum\nolimits_{a} \Tr(\lap_{a}\bro)\, R_{\gamma}(\clb;\taug_{a})
\, . \label{avbr}
\end{equation}

Of course, the first sum in the right-hand side of (\ref{avbr})
vanishes. Measuring $\amh$ in its eigenstate leads to a
deterministic probability distribution, whence
$R_{\alpha}(\cla;\taug_{a})=0$ for all $a$. It is for this reason
that only the left-hand side of (\ref{avbr}) is used in studies of
uncertainties in successive measurements of finite-dimensional
observables. In a similar
manner, we can rewrite (\ref{rescp}) and (\ref{avbr}) with the use
of Tsallis' entropies. For $\alpha=\gamma=1$, the quantity
(\ref{avbr}) becomes the Shannon entropy averaged over all $a$.
The authors of \cite{bfs2014} utilized the latter as a measure of
uncertainties in successive measurements. Uncertainty relations
for successive projective measurements in terms of R\'{e}nyi's
entropies were analyzed in \cite{zzhang14}. Formally, the sums
involved in (\ref{avbr}) are similar to one of several existing
definitions of conditional R\'{e}nyi's entropy. In more detail,
these definitions are discussed \cite{tma12}. The simplest of them
just leads to expressions of the form (\ref{avbr}). Moreover, the
two kinds of conditional Tsallis entropy are known in the
literature \cite{sf06,rastkyb}. More properties of
generalized conditional entropies are discussed in
\cite{rastit}.

Let us proceed to exact formulations for successive measurements
of position and momentum. One cannot provide
states in which the measurement of position or momentum
gives exactly one particular value. Instead, we deal with well
localized states of finite or even small scales. Following
\cite{cancon16}, the right-hand side of (\ref{avbr}) will be used
in extending the second scenario to the position-momentum case in
the presence of a minimal length. Suppose that the first
applied measurement aims to measure momentum. The authors of
\cite{paban13} mentioned how the post-first-measurement state
should be posed. In our notation, we write
\begin{equation}
\Phi_{\clm}(\bro)=
\int_{-\infty}^{+\infty}\xdif\zeta\, \mmh(\zeta)\bro\mmh(\zeta)^{\dagger}
\, . \label{fspm1}
\end{equation}

This expression replaces the formula (\ref{fsmpm}) suitable for
observables with a purely discrete spectrum. The following
important fact should be pointed out. If we again measure
momentum, but now with the state (\ref{fspm1}), then it will
result in the same probability distribution function. It can be
derived from (\ref{mmzt}) that
\begin{equation}
\langle{k}|\bro|k\rangle=\langle{k}|\Phi_{\clm}(\bro)|k\rangle
\, , \qquad
U_{\bro}(\zeta)=U_{\Phi_{\clm}(\bro)}(\zeta)
\, . \label{upur}
\end{equation}

Such relations may be interpreted as a mild version of the
repeatability concept. For strictly positive $\alpha\neq1$, the
R\'{e}nyi $\alpha$-entropy $R_{\alpha}(\clm;\bro)$ is given by
substituting $U_{\bro}(\zeta)$ into (\ref{reddf}). The standard
differential entropy $H_{1}(\clm;\bro)$ can be obtained within the
limit $\alpha\to1$. Also, the R\'{e}nyi $\alpha$-entropy
$R_{\alpha}\bigl(\pq_{\clm}^{(\delta)};\bro\bigr)$ is defined by
(\ref{rpdf}) by substituting probabilities defined through a
discretization of the $\zeta$-axis. When the first measurement is
described by the set $\cln$, the post-first-measurement state is
specified as
\begin{equation}
\Phi_{\cln}(\bro)=
\int_{-\infty}^{+\infty}\xdif\xi\, \nnh(\xi)\bro\nnh(\xi)^{\dagger}
\, . \label{fspn1}
\end{equation}

Let $\bro$ denote the state right before the sequence of
successive measurements. In the first scenario of successive
measurements, we will characterize uncertainties by entropic
quantities of the form
\begin{equation}
R_{\alpha}(\clm;\bro)+R_{\gamma}\bigl(\cln;\Phi_{\clm}(\bro)\bigr)
\, , \qquad
R_{\alpha}\bigl(\clm;\Phi_{\cln}(\bro)\bigr)+R_{\gamma}(\cln;\bro)
\, . \label{refimp}
\end{equation}

The former of the two sums concerns the case in which momentum is
measured. Another useful approach is to calculate
entropies with binning. For instance, sampling of the function
(\ref{bigu}) into bins between marks $\zeta_{j}$ gives a discrete
probability distribution $\pq_{\clm}^{(\delta)}$. In the second
measurement, entropies can be taken with binning between some
marks $\xi_{k}$. By $\pq_{\cln}^{(\delta)}$, we mean the
corresponding probability distribution. This approach leads to the
characteristic quantities
\begin{equation}
R_{\alpha}\bigl(\pq_{\clm}^{(\delta)};\bro\bigr)
+R_{\gamma}\bigl(\pq_{\cln}^{(\delta)};\Phi_{\clm}(\bro)\bigr)
\, , \qquad
R_{\alpha}\bigl(\pq_{\clm}^{(\delta)};\Phi_{\cln}(\bro)\bigr)
+R_{\gamma}\bigl(\pq_{\cln}^{(\delta)};\bro\bigr)
\, . \label{rebimp}
\end{equation}

In a similar manner, we formulate entropic measures of the Tsallis
type. As was already mentioned, such entropies will be taken only
with binning.

The second scenario of successive measurements prescribes that
each actual result of the first measurement should be retained.
Assuming $U_{\bro}(\zeta)\neq0$ in the corresponding domain, we
now consider the normalized output state
\begin{equation}
\vbro(\zeta)=U_{\bro}(\zeta)^{-1}\mmh(\zeta)\bro\mmh(\zeta)^{\dagger}
\, . \label{vbrz}
\end{equation}

Each $\vbro(\zeta)$ is used as one of possible pre-measurement
states in the second measurement. Similarly to (\ref{avbr}), we
then consider the quantity
\begin{equation}
\int_{-\infty}^{+\infty} R_{\alpha}\bigl(\clm;\vbro(\zeta)\bigr) U_{\bro}(\zeta)\,\xdif\zeta +
\int_{-\infty}^{+\infty} R_{\gamma}\bigl(\cln;\vbro(\zeta)\bigr) U_{\bro}(\zeta)\,\xdif\zeta
\, . \label{secnob}
\end{equation}

When position is measured by the first, particular outputs are of
the form
\begin{equation}
\bsg(\xi)=W_{\bro}(\xi)^{-1}\nnh(\xi)\bro\nnh(\xi)^{\dagger}
\, . \label{bsrz}
\end{equation}

To describe the amount of uncertainty here, we rewrite
(\ref{secnob}) with $\bsg(\xi)$ instead of $\vbro(\zeta)$ and
$W_{\bro}(\xi)$ instead of $U_{\bro}(\zeta)$. We will also utilize
entropic uncertainty relations with binning. Here, one replaces
(\ref{secnob}) with
\begin{equation}
\int_{-\infty}^{+\infty} R_{\alpha}\bigl(\pq_{\clm}^{(\delta)};\vbro(\zeta)\bigr) U_{\bro}(\zeta)\,\xdif\zeta +
\int_{-\infty}^{+\infty} R_{\gamma}\bigl(\pq_{\cln}^{(\delta)};\vbro(\zeta)\bigr) U_{\bro}(\zeta)\,\xdif\zeta
\, , \label{secbb}
\end{equation}
and similarly with the Tsallis entropies. Quantities of the form
(\ref{secbb}) concern successive measurements, in which position
is measured after momentum. When position is measured by the
first, we rewrite such expressions with $\bsg(\xi)$ and
$W_{\bro}(\xi)$. In the paper \cite{cancon16}, the above treatment
of successive measurements was considered for general canonically
conjugate operators. This approach to the concept of canonical
conjugacy is based on the Pegg--Barnett formalism \cite{pvb90}.
The Pegg--Barnett formalism was originally proposed to explain a
Hermitian phase operator \cite{PB89,BP89}. Entropic uncertainty
relations on the base of this formalism were examined in
\cite{abe92,gvb95,rastnum12}.

\section{Main results}\label{sec4}

In this section, we shall formulate entropic uncertainty relations
for successive measurements within the GUP case. For this case,
preparation uncertainty relations with a correction term were
derived in \cite{gegen17}. For the convenience of further
calculations, the prepared pre-measurement state will be denoted
by $\bomg$. Due to \cite{gegen17}, we have
\begin{equation}
H_{1}(\clm;\bomg)+H_{1}(\cln;\bomg)\geq
H_{1}(\clk;\bomg)+H_{1}(\clx;\bomg)\geq
\ln(e\pi)
+\bigl\langle\ln(1+\beta\skx^{2})\bigr\rangle_{\bomg}
\, . \label{bbmn1}
\end{equation}

The well-known bound
$\ln(e\pi)$ corresponds to the entropic uncertainty relation of
Beckner \cite{beck} and Bia{\l}ynicki-Birula and Mycielski
\cite{birula1}. The second term in the right-hand side of
(\ref{bbmn1}) reflects the fact that the legitimate momentum of
the commutation relation (\ref{gcomr}) is given by $\hbar\skx$.
Here, the wavenumber operator $\sqx$ plays an auxiliary role. Note
that this correction term depends on the pre-measurement state. As
some numerical results in \cite{kaw17} later showed, the
presented correction is sufficiently tight. It is similar to the
correction term obtained in the Robertson formulation (\ref{robr}).
However, the inequality (\ref{bbmn1}) is a preparation uncertainty
relation.

Suppose now that we measure momentum by the first and position by
the second. In the first scenario, the pre-measurement state
$\bro$ leads to the post-first-measurement state
$\Phi_{\clm}(\bro)$. Due to (\ref{upur}), we immediately write
\begin{equation}
H_{1}(\clm;\bro)=H_{1}\bigl(\clm;\Phi_{\clm}(\bro)\bigr)
\, , \qquad
\bigl\langle\ln(1+\beta\skx^{2})\bigr\rangle_{\bro}=
\bigl\langle\ln(1+\beta\skx^{2})\bigr\rangle_{\Phi_{\clm}(\bro)}
\, . \label{twocs}
\end{equation}

Substituting $\bomg=\Phi_{\clm}(\bro)$ into (\ref{bbmn1}) and
using (\ref{twocs}), we easily get
\begin{equation}
H_{1}(\clm;\bro)+H_{1}\bigl(\cln;\Phi_{\clm}(\bro)\bigr)\geq\ln(e\pi)
+\bigl\langle\ln(1+\beta\skx^{2})\bigr\rangle_{\bro}
\, . \label{h1fmp}
\end{equation}

This is an entropic uncertainty relation in the first scenario of
successive measurements such that momentum is measured by the
first. The corresponding lower bound is the same as in the
preparation scenario. It is not the case, when we measure position
by the first and momentum by the second. Putting
$\bomg=\Phi_{\cln}(\bro)$ into (\ref{bbmn1}) finally gives
\begin{equation}
H_{1}(\cln;\bro)+H_{1}\bigl(\clm;\Phi_{\cln}(\bro)\bigr)\geq\ln(e\pi)
+\bigl\langle\ln(1+\beta\skx^{2})\bigr\rangle_{\Phi_{\cln}(\bro)}
\, . \label{h1fpm}
\end{equation}

The correction term in the right-hand side of (\ref{h1fpm}) is
similar in form but should be calculated with the
post-first-measurement state $\Phi_{\cln}(\bro)$. Taking
$\beta=0$, the above entropic bounds for successive measurements
do not differ from the bound in the preparation scenario. Here, we
see a manifestation of the deformed commutation relation
(\ref{gcomr}). The latter disturbs a certain symmetry between
position and momentum.

Let us proceed to the second scenario of successive measurements.
Suppose again that momentum is measured by the first. Substituting
$\bomg=\vbro(\zeta)$ into (\ref{bbmn1}), we multiply it by
$U_{\bro}(\zeta)$ and then integrate with respect to $\zeta$. This
results in the inequality
\begin{equation}
\int_{-\infty}^{+\infty} H_{1}\bigl(\clm;\vbro(\zeta)\bigr) U_{\bro}(\zeta)\,\xdif\zeta +
\int_{-\infty}^{+\infty} H_{1}\bigl(\cln;\vbro(\zeta)\bigr) U_{\bro}(\zeta)\,\xdif\zeta
\geq
\ln(e\pi)+
\int_{-\infty}^{+\infty} \bigl\langle\ln(1+\beta\skx^{2})\bigr\rangle_{\vbro(\zeta)}\,U_{\bro}(\zeta)\,\xdif\zeta
\, . \label{secmp0}
\end{equation}

Using (\ref{vbrz}), the second term in the right-hand side of
(\ref{secmp0}) can be simplified, viz.,
\begin{align}
\int_{-\infty}^{+\infty}\xdif\zeta\,U_{\bro}(\zeta)\int_{-\infty}^{+\infty}\xdif{k}\,\ln(1+\beta{k}^{2})\,\langle{k}|\vbro(\zeta)|k\rangle
&=\int_{-\infty}^{+\infty}\xdif\zeta\,\int_{-\infty}^{+\infty}\xdif{k}\,\ln(1+\beta{k}^{2})\,\langle{k}|\mmh(\zeta)\bro\mmh(\zeta)^{\dagger}|k\rangle
\nonumber\\
&=\int_{-\infty}^{+\infty}\xdif{k}\,\ln(1+\beta{k}^{2})\,\langle{k}|\bro|k\rangle\int_{-\infty}^{+\infty}\xdif\zeta\,|f(\zeta-k)|^{2}
\, . \label{secnor}
\end{align}

In the right-hand side of (\ref{secnor}), the last integral with
respect to $\zeta$ is equal to $1$. For the second scenario of
successive measurements, we obtain
\begin{equation}
\int_{-\infty}^{+\infty} H_{1}\bigl(\clm;\vbro(\zeta)\bigr) U_{\bro}(\zeta)\,\xdif\zeta +
\int_{-\infty}^{+\infty} H_{1}\bigl(\cln;\vbro(\zeta)\bigr) U_{\bro}(\zeta)\,\xdif\zeta
\geq
\ln(e\pi)+\bigl\langle\ln(1+\beta\skx^{2})\bigr\rangle_{\bro}
\, . \label{secmp1}
\end{equation}

So, entropic uncertainty relations (\ref{h1fmp}) and
(\ref{secmp1}) are obtained with the same lower bound calculated
with the pre-measurement state. Let us consider the case when
position is measured by the first. Substituting
$\bomg=\bsg(\xi)$ into (\ref{bbmn1}), we multiply it by
$W_{\bro}(\xi)$ and integrate with respect to $\xi$, whence
\begin{equation}
\int_{-\infty}^{+\infty} H_{1}\bigl(\clm;\bsg(\xi)\bigr) W_{\bro}(\xi)\,\xdif\xi +
\int_{-\infty}^{+\infty} H_{1}\bigl(\cln;\bsg(\xi)\bigr) W_{\bro}(\xi)\,\xdif\xi
\geq
\ln(e\pi)+
\int_{-\infty}^{+\infty} \bigl\langle\ln(1+\beta\skx^{2})\bigr\rangle_{\bsg(\xi)}\,W_{\bro}(\xi)\,\xdif\xi
\, . \label{secpm0}
\end{equation}

In the right-hand side of (\ref{secpm0}), the second integral is a
correction term averaged over particular outputs $\bsg(\xi)$. In
general, an expression for this term cannot be simplified without
additional assumptions. We have already seen how the relation
(\ref{h1fpm}) differs from (\ref{h1fmp}). The formula
(\ref{secpm0}) differs from (\ref{secmp1}) in a similar vein. In
the presence of a minimal length, the preparation uncertainty
relation (\ref{bbmn1}) remains valid for successive measurements,
when momentum is measured by the first. Otherwise, it should be
reformulated.

Entropic uncertainty relations with binning can be treated in a
similar manner. Using some discretization of axes, we take into
account sufficiently typical setup. This approach also leads to
entropic functions with only positive values. In contrast,
differential entropies can generally have arbitrary signs. In the
case of momentum measurements, values $\zeta_{i}$ denote the ends
of intervals $\delta\zeta_{i}=\zeta_{i+1}-\zeta_{i}$. For the
prepared state $\bomg$, we deal with probabilities
\begin{equation}
p_{i}^{(\delta)}:=\int\nolimits_{\zeta_{i}}^{\zeta_{i+1}}
 U_{\bomg}(\zeta)\,\xdif\zeta
\, , \label{pduz}
\end{equation}
which form the discrete distribution $\pq_{\clm}^{(\delta)}$.
Using (\ref{pduz}), one calculates the Shannon entropy
$H_{1}(\pq_{\clm}^{(\delta)};\bomg)$. In a similar way, we
discretize the $\xi$-axis into bins
$\delta\xi_{j}=\xi_{j+1}-\xi_{j}$ with the resulting distribution
$\pq_{\cln}^{(\delta)}$. It can be shown that
\begin{equation}
H_{1}\bigl(\pq_{\clm}^{(\delta)};\bomg\bigr)+H_{1}\bigl(\pq_{\cln}^{(\delta)};\bomg\bigr)\geq
\ln\!\left(\frac{e\pi}{\delta\zeta\,\delta\xi}\right)
+\bigl\langle\ln(1+\beta\skx^{2})\bigr\rangle_{\bomg}
\, , \label{bin1om}
\end{equation}
where $\delta\zeta=\max\delta\zeta_{i}$ and
$\delta\xi=\max\delta\xi_{j}$. The formula (\ref{bin1om}) gives a
preparation uncertainty relation with binning. It involves the
same correction term due to the existence of a minimal length. To
pose entropic uncertainty relations in the first scenario of
successive measurements, we again use reasons that have led to
(\ref{h1fmp}) and (\ref{h1fpm}). Finally, one gets
\begin{align}
H_{1}\bigl(\pq_{\clm}^{(\delta)};\bro\bigr)+H_{1}\bigl(\pq_{\cln}^{(\delta)};\Phi_{\clm}(\bro)\bigr)&\geq
\ln\!\left(\frac{e\pi}{\delta\zeta\,\delta\xi}\right)
+\bigl\langle\ln(1+\beta\skx^{2})\bigr\rangle_{\bro}
\, , \label{h1dmp}\\
H_{1}\bigl(\pq_{\cln}^{(\delta)};\bro\bigr)+H_{1}\bigl(\pq_{\clm}^{(\delta)};\Phi_{\cln}(\bro)\bigr)&\geq
\ln\!\left(\frac{e\pi}{\delta\xi\,\delta\zeta}\right)
+\bigl\langle\ln(1+\beta\skx^{2})\bigr\rangle_{\Phi_{\cln}(\bro)}
\, . \label{h1dpm}
\end{align}

In the second scenario of successive measurements, entropic
uncertainty relations with binning are obtained in the form
\begin{align}
&\int_{-\infty}^{+\infty} H_{1}\bigl(\pq_{\clm}^{(\delta)};\vbro(\zeta)\bigr) U_{\bro}(\zeta)\,\xdif\zeta
+\int_{-\infty}^{+\infty} H_{1}\bigl(\pq_{\cln}^{(\delta)};\vbro(\zeta)\bigr) U_{\bro}(\zeta)\,\xdif\zeta
\geq
\ln\!\left(\frac{e\pi}{\delta\zeta\,\delta\xi}\right)
+\bigl\langle\ln(1+\beta\skx^{2})\bigr\rangle_{\bro}
\, , \label{bsecmp}\\
&\int_{-\infty}^{+\infty} H_{1}\bigl(\pq_{\cln}^{(\delta)};\bsg(\xi)\bigr) W_{\bro}(\xi)\,\xdif\xi
+\int_{-\infty}^{+\infty} H_{1}\bigl(\pq_{\clm}^{(\delta)};\bsg(\xi)\bigr) W_{\bro}(\xi)\,\xdif\xi
\geq
\ln\!\left(\frac{e\pi}{\delta\xi\,\delta\zeta}\right)
+\int_{-\infty}^{+\infty} \bigl\langle\ln(1+\beta\skx^{2})\bigr\rangle_{\bsg(\xi)}W_{\bro}(\xi)\,\xdif\xi
\, . \label{bsecpm}
\end{align}

In the presence of a minimal length, distinctions of (\ref{h1dpm})
and (\ref{bsecpm}) from the corresponding preparation relations
are concentrated in correction terms. In effect, these terms are
not state-independent. On the other hand, entropic bounds of
preparation uncertainty relations remain valid, when momentum is
measured by the first. The author of \cite{gegen17} also reported on state-independent
entropic uncertainty relations in the presence of a minimal
length. Such relations were posed in terms of the R\'{e}nyi and
Tsallis entropies with binning. An alteration of statistics due to
a finite resolution of the measurements is also taken into
account. When acceptance functions of measurement apparatuses are
sufficiently spread, they lead to an increase of entropic lower
bounds. To pose uncertainty relations formally, we introduce the
following quantity \cite{gegen17}, 
\begin{equation}
S_{f}:=\underset{\zeta}{\sup}\int_{-\infty}^{+\infty}\frac{|f(\zeta-k)|^{2}}{1+\beta{k}^{2}}\>\xdif{k}
\, , \label{sfedf}
\end{equation}
where the acceptance function $\zeta\mapsto{f}(\zeta)$ corresponds
to momentum measurements. Let $\bomg$ represent the prepared state.
As was shown in \cite{gegen17}, the existence of a minimal length
leads to preparation uncertainty relations of the form
\begin{equation}
R_{\alpha}(\clm;\bomg)+R_{\gamma}(\cln;\bomg)
\geq\ln\!\left(\frac{\varkappa\pi}{S_{f}}\right)
 . \label{remng}
\end{equation}

Here, positive entropic parameters obey $1/\alpha+1/\gamma=2$ and
\begin{equation}
\varkappa^{2}=\alpha^{1/(\alpha-1)}\gamma^{1/(\gamma-1)}
\, . \label{mpmpf}
\end{equation}

In the limit $\alpha\to1$, the parameter $\varkappa$ becomes equal
to $e$. When $\beta=0$, we clearly have $S_{f}=1$, so that the
right-hand side of (\ref{remng}) reduces to $\ln(\varkappa\pi)$.
The latter is a known entropic bound for the case of usual
position and momentum. For $\beta>0$ and physically reasonable
acceptance functions, we obtain an improved lower bound due to
$S_{f}<1$. It is important that the quantity (\ref{sfedf}) depends
only on $\beta$ and the actual acceptance function in momentum
measurements. Preparation entropic uncertainty relations with
binning are posed as follows \cite{gegen17}. Let probability
density functions $U_{\bomg}(\zeta)$ and $W_{\bomg}(\xi)$ be
sampled into discrete probability distributions. Then the
corresponding R\'{e}nyi and Tsallis entropies satisfy
\begin{align}
R_{\alpha}\bigl(\pq_{\clm}^{(\delta)};\bomg\bigr)+R_{\gamma}\bigl(\pq_{\cln}^{(\delta)};\bomg\bigr)
&\geq\ln\!\left(\frac{\varkappa\pi}{S_{f}\,\delta\zeta\,\delta\xi}\right)
 , \label{rebng}\\
H_{\alpha}\bigl(\pq_{\clm}^{(\delta)};\bomg\bigr)+H_{\gamma}\bigl(\pq_{\cln}^{(\delta)};\bomg\bigr)
&\geq\ln_{\nu}\!\left(\frac{\varkappa\pi}{S_{f}\,\delta\zeta\,\delta\xi}\right)
 , \label{tsbng}
\end{align}
where $1/\alpha+1/\gamma=2$ and $\nu=\max\{\alpha,\gamma\}$.

Due to equalities of the form (\ref{upur}), the preparation
uncertainty relations (\ref{remng}), (\ref{rebng}), and
(\ref{tsbng}) are immediately converted into relations for
successive measurements. In the first scenario, we obtain
\begin{equation}
R_{\alpha}(\clm;\bro)+R_{\gamma}\bigl(\cln;\Phi_{\clm}(\bro)\bigr)
\geq\ln\!\left(\frac{\varkappa\pi}{S_{f}}\right)
 , \label{remngsm}
\end{equation}
where $1/\alpha+1/\gamma=2$ and the momentum measurement is
assumed to be made by the first. When position is measured by the
first, we replace $\bro$ with $\Phi_{\cln}(\bro)$ and
$\Phi_{\clm}(\bro)$ with $\bro$ in the left-hand side of
(\ref{remngsm}). For $1/\alpha+1/\gamma=2$ and
$\nu=\max\{\alpha,\gamma\}$, entropic uncertainty relations with
binning are written as
\begin{align}
R_{\alpha}\bigl(\pq_{\clm}^{(\delta)};\bro\bigr)+R_{\gamma}\bigl(\pq_{\cln}^{(\delta)};\Phi_{\clm}(\bro)\bigr)
&\geq\ln\!\left(\frac{\varkappa\pi}{S_{f}\,\delta\zeta\,\delta\xi}\right)
 , \label{rebngsm}\\
H_{\alpha}\bigl(\pq_{\clm}^{(\delta)};\bro\bigr)+H_{\gamma}\bigl(\pq_{\cln}^{(\delta)};\Phi_{\clm}(\bro)\bigr)
&\geq\ln_{\nu}\!\left(\frac{\varkappa\pi}{S_{f}\,\delta\zeta\,\delta\xi}\right)
 . \label{tsbngsm}
\end{align}

The same entropic lower bounds hold, when position is measured by
the first. We refrain from presenting the details here. In the
second scenario of successive measurements, one immediately gets
\begin{equation}
\int_{-\infty}^{+\infty} R_{\alpha}\bigl(\clm;\vbro(\zeta)\bigr) U_{\bro}(\zeta)\,\xdif\zeta+
\int_{-\infty}^{+\infty} R_{\gamma}\bigl(\cln;\vbro(\zeta)\bigr) U_{\bro}(\zeta)\,\xdif\zeta
\geq\ln\!\left(\frac{\varkappa\pi}{S_{f}}\right)
 , \label{remngms}
\end{equation}
where $1/\alpha+1/\gamma=2$ and the momentum measurement is
assumed to be made by the first. Replacing $\vbro(\zeta)$ with
$\bsg(\xi)$ and $U_{\bro}(\zeta)$ with $W_{\bro}(\xi)$, we resolve
the case when position is measured by the first. For
$1/\alpha+1/\gamma=2$ and $\nu=\max\{\alpha,\gamma\}$, entropic
uncertainty relations with binning are expressed as
\begin{align}
\int_{-\infty}^{+\infty} R_{\alpha}\bigl(\pq_{\clm}^{(\delta)};\vbro(\zeta)\bigr) U_{\bro}(\zeta)\,\xdif\zeta+
\int_{-\infty}^{+\infty} R_{\gamma}\bigl(\pq_{\cln}^{(\delta)};\vbro(\zeta)\bigr) U_{\bro}(\zeta)\,\xdif\zeta
&\geq\ln\!\left(\frac{\varkappa\pi}{S_{f}\,\delta\zeta\,\delta\xi}\right)
 , \label{rebngms}\\
\int_{-\infty}^{+\infty} H_{\alpha}\bigl(\pq_{\clm}^{(\delta)};\vbro(\zeta)\bigr) U_{\bro}(\zeta)\,\xdif\zeta+
\int_{-\infty}^{+\infty} H_{\gamma}\bigl(\pq_{\cln}^{(\delta)};\vbro(\zeta)\bigr) U_{\bro}(\zeta)\,\xdif\zeta
&\geq\ln_{\nu}\!\left(\frac{\varkappa\pi}{S_{f}\,\delta\zeta\,\delta\xi}\right)
 . \label{tsbngms}
\end{align}

When position is measured by the first, we merely replace here
$\vbro(\zeta)$ with $\bsg(\xi)$ and $U_{\bro}(\zeta)$ with
$W_{\bro}(\xi)$. Thus, state-independent entropic lower bounds of
preparation uncertainty relation remain valid for scenarios with
successive measurements. The existence of a minimal length is
taken into account due to the quantity (\ref{sfedf}). In the case
$\alpha=\gamma=1$, the above relations are expressed via the
Shannon entropies. We have also obtained state-dependent entropic
uncertainty relations such as (\ref{h1fpm}), (\ref{secpm0}),
(\ref{h1dpm}), and (\ref{bsecpm}). Their formulation differ from
preparation uncertainty relations since they depend on the quantum state
immediately following the first measurement.

\section{Conclusions}\label{sec5}

We have formulated entropic uncertainty relations for successive
measurements in the presence of a minimal length. The presented
formulation is explicitly dependent on the order of the
measurements, though the bounds themselves may not be optimal. The
problem of a minimal observable length is related to efforts to
describe gravitation at the quantum level. In effect, the
generalized uncertainty principle restricts the space of
acceptable wave packets. Scenarios with successive measurements
are interesting for several reasons. The traditional scenario of
preparation uncertainty relations is insufficient even from the
viewpoint of Heisenberg's thought experiment \cite{heisenberg}.
Successive measurements of position and momentum cannot be treated
as projective even within ordinary quantum mechanics. The GUP case
implies additional limitation for a spatial width of the
acceptance function in position measurements. Thus, entropic
measures of uncertainty should be formulated differently from the
finite-dimensional case. One of distinctions concerns a proper
form of the state immediately following the first measurement. The
post-first-measurement state was chosen according to the two
possible scenarios. Uncertainty relations in terms of Shannon
entropies contain a state-dependent correction term. Hence,
entropic lower bounds for successive measurements generally differ
from lower bounds involved into preparation uncertainty relations.
We also formulated state-independent uncertainty bounds in terms
of R\'{e}nyi entropies and, with binning, in terms of Tsallis
entropies. In the presence of a minimal length, state-independent
entropic lower bounds of preparation uncertainty relation remain
valid for scenarios with successive measurements. When acceptance
functions of measurement apparatuses are sufficiently spread, the
existing entropic lower bounds are improved.


\begin{thebibliography}{00}

\bibitem{heisenberg}
Heisenberg, W. \"{U}ber den anschaulichen inhalt der quanten theoretischen kinematik und mechanik. {\it Z. Phys.} {\bf 1927}, {\it 43}, 172--198.

\bibitem{kennard}
Kennard, E.H. Zur quantenmechanik einfacher bewegungstypen. {\it Z. Phys.} {\bf 1927}, {\it 44}, 326--352.

\bibitem{robert}
Robertson, H.P. The uncertainty principle. {\it Phys. Rev.} {\bf 1929}, {\it 34}, 163--164.

\bibitem{rozp17}
Rozp\c{e}dek, F.; Kaniewski, J.; Coles, P.J.; Wehner, S. Quantum preparation uncertainty and lack of information. {\it New J. Phys.} {\bf 2017}, {\it 19}, 023038.

\bibitem{ckw14}
Coles, P.J.; Kaniewski, J.; Wehner, S. Equivalence of wave-particle duality to entropic uncertainty. {\it Nat. Commun.} {\bf 2014}, {\it 5}, 5814.

\bibitem{ww10}
Wehner, S.; Winter, A. Entropic uncertainty relations -- a survey, {\it New J. Phys.} {\bf 2010}, {\it 12}, 025009.

\bibitem{brud11}
Bia{\l}ynicki-Birula, I.; Rudnicki, {\L}. Entropic uncertainty
relations in quantum physics. In {\it Statistical Complexity.}
Springer, Berlin {\bf 2011}, 1--34.

\bibitem{cbtw17}
Coles, P.J.; Berta, M.; Tomamichel, M.; Wehner, S. Entropic
uncertainty relations and their applications, {\it Rev. Mod.
Phys.} {\bf 2017}, {\it 89}, 015002.

\bibitem{renner10}
Berta, M.; Christandl, M.; Colbeck, R.; Renes, J.M.; Renner, R.
The uncertainty principle in the presence of quantum memory. {\it
Nature Phys.} {\bf 2010}, {\it 6}, 659--662.

\bibitem{renner11}
Tomamichel, M.; Renner, R. Uncertainty relation for smooth entropies. {\it Phys. Rev. Lett.} {\bf 2011}, {\it 106}, 110506.

\bibitem{nbw12}
Ng, N.H.Y.; Berta, M.; Wehner, S. Min-entropy uncertainty relation for finite-size cryptography. {\it Phys. Rev. A} {\bf 2012}, {\it 86}, 042315.

\bibitem{furrer14}
Furrer, F. Reverse-reconciliation continuous-variable quantum key
distribution based on the uncertainty principle. {\it Phys. Rev.
A} {\bf 2014}, {\it 90}, 042325.

\bibitem{lfei18}
Li, J.; Fei, S.-M. Uncertainty relation based on Wigner--Yanase--Dyson skew information with quantum memory. {\it Entropy} {\bf 2018}, {\it 20}, 132.

\bibitem{mdsrin03}
Srinivas, M.D. Optimal entropic uncertainty relation for
successive measurements in quantum information theory. {\it
Pramana--J. Phys.} {\bf 2003}, {\it 60}, 1137--1152.

\bibitem{paban13}
Distler, J.; Paban, S. Uncertainties in successive measurements. {\it Phys. Rev. A} {\bf 2013}, {\it 87}, 062112

\bibitem{bfs2014}
Baek, K.; Farrow, T.; Son, W. Optimized entropic uncertainty for successive projective measurements. {\it Phys. Rev. A} {\bf 2014}, {\it 89}, 032108.

\bibitem{zzhang14}
Zhang, J.; Zhang, Y.; Yu, C.-S. R\'{e}nyi entropy uncertainty
relation for successive projective measurements. {\it Quantum Inf.
Process.} {\bf 2015}, {\it 14}, 2239--2253.

\bibitem{bs2016}
Baek, K.; Son, W. Entropic uncertainty relations for successive generalized measurements. {\it Mathematics} {\bf 2016}, {\it 4}, 41.

\bibitem{wineland13}
Wineland, D. Superposition, entanglement, and raising Schr\"{o}dinger's cat. {\it Ann. Phys. (Berlin)} {\bf 2013}, {\it 525}, 739--752.

\bibitem{haroche13}
Haroche, S. Controlling photons in a box and exploring the quantum to classical boundary. {\it Ann. Phys. (Berlin)} {\bf 2013}, {\it 525}, 753--776.

\bibitem{blw13}
Busch, P.; Lahti, P.; Werner, R.F. Proof of Heisenberg’s error-disturbance relation. {\it Phys. Rev. Lett.} {\bf 2013}, {\it 111}, 160405.

\bibitem{ozawa04}
Ozawa, M. Uncertainty relations for noise and disturbance in generalized quantum measurements. {\it Ann. Phys.} {\bf 2004}, {\it 311}, 350--416.

\bibitem{gisin18}
Gisin, N. Why Bohmian Mechanics? One- and two-time position
measurements, Bell inequalities, philosophy, and physics. {\it
Entropy} {\bf 2018}, {\it 20}, 105.

\bibitem{hossen13}
Hossenfelder, S. Minimal length scale scenarios for quantum gravity. {\it Living Rev. Relativity} {\bf 2013}, {\it 16}, 2.

\bibitem{rovelli04}
Rovelli, C. Quantum Gravity. {\it Cambridge University Press, Cambridge} {\bf 2004}.

\bibitem{amati89}
Amati, D.; Ciafaloni, M.; Veneziano, G. Can spacetime be probed below the string size? {\it Phys. Lett. B} {\bf 1989}, {\it 216}, 41--47.

\bibitem{scard99}
Scardigli, F. Generalized uncertainty principle in quantum gravity
from micro-black hole gedanken experiment. {\it Phys. Lett. B}
{\bf 1999}, {\it 452}, 39--44.

\bibitem{bombi08}
Bambi, C. A revision of the generalized uncertainty principle. {\it Class. Quantum Grav.} {\bf 2008}, {\it 25}, 105003.

\bibitem{td2015}
Tawfik, A.N.; Diab, A.M. A review of the generalized uncertainty principle. {\it Rep. Prog. Phys.} {\bf 2015}, {\it 78}, 126001.

\bibitem{ellis98}
Amelino-Camelia, G.; Ellis, J.; Mavromatos, N.E.; Nanopoulos,
D.V.; Sarkar, S. Tests of quantum gravity from observations of
$\gamma$-ray bursts. {\it Nature} {\bf 1998}, {\it 393}, 763--765.

\bibitem{piran07}
Jacob, U.; Piran, T. Neutrinos from gamma-ray bursts as a tool to
explore quantum-gravity-induced Lorentz violation. {\it Nature
Phys.} {\bf 2007}, {\it 3}, 87--90.

\bibitem{brukner12}
Pikovski, I.; Vanner, M.R.; Aspelmeyer, M.; Kim, M.; Brukner,
\v{C}. Probing Planck-scale physics with quantum optics. {\it
Nature Phys.} {\bf 2012}, {\it 8}, 393--397.

\bibitem{ffm13}
Marin, F.; Marino, F.; Bonaldi, M.; Cerdonio, M.; Conti, L.;
Falferi, P.; Mezzena, R.; Ortolan, A.; Prodi, G.A.; Taffarello,
L.; Vedovato, G.; Vinante, A.; Zendri, J.-P. Gravitational bar
detectors set limits to Planck-scale physics on macroscopic
variables. {\it Nature Phys.} {\bf 2013}, {\it 9}, 71--73.

\bibitem{tawfik13}
Tawfik, A. Impacts of generalized uncertainty principle on black
hole thermodynamics and Salecker--Wigner inequalities. {\it JCAP}
{\bf 2013}, {\it 07(2013)}, 040.

\bibitem{deys13}
Dey, S.; Fring, A.; Khantoul, B. Hermitian versus non-Hermitian
representations for minimal length uncertainty relations. {\it J.
Phys. A: Math. Theor.} {\bf 2013}, {\it 46}, 335304.

\bibitem{diab14}
Tawfik, A.; Diab, A. Generalized uncertainty principle: Approaches and applications. {\it Int. J. Mod. Phys. A} {\bf 2014}, {\it 23}, 1430025.

\bibitem{faizalm15}
Faizal, M.; Majumder, B. Incorporation of generalized uncertainty principle into Lifshitz field theories. {\it Ann. Phys.} {\bf 2015}, {\it 357}, 49--58.

\bibitem{mfza16}
Masood, S.; Faizal, M.; Zaz, Z.; Ali, A.F.; Raza, J.; Shah, M.B.
The most general form of deformation of the Heisenberg algebra
from the generalized uncertainty principle. {\it Phys. Lett. B}
{\bf 2016}, {\it 763}, 218--227.

\bibitem{kempf95}
Kempf, A.; Mangano, G.; Mann, R.B. Hilbert space representation of
the minitial length uncertainty relation. {\it Phys. Rev. D} {\bf
1995}, {\it 52}, 1108--1118.

\bibitem{pedram12}
Pedram, P. New approach to nonperturbative quantum mechanics with minimal length uncertainty. {\it Phys. Rev. D} {\bf 2012}, {\it 85}, 024016.

\bibitem{pedram16}
Pedram, P. The minimal length and the Shannon entropic uncertainty relation. {\it Adv. High Energy Phys.} {\bf 2016}, {\it 2016}, 5101389.

\bibitem{beck}
Beckner, W. Inequalities in Fourier analysis. {\it Ann. Math.} {\bf 1975}, {\it 102}, 159--182.

\bibitem{birula1}
Bia{\l}ynicki-Birula, I.; Mycielski, J. Uncertainty relations for
information entropy in wave mechanics. {\it Commun. Math. Phys.}
{\bf 1975}, {\it 44}, 129--132.

\bibitem{hirs}
Hirschman, I.I. A note on entropy. {\it Amer. J. Math.} {\bf 1957}, {\it 79}, 152--156.

\bibitem{deutsch}
Deutsch, D. Uncertainty in quantum measurements. {\it Phys. Rev. Lett.} {\bf 1983}, {\it 50}, 631--633.

\bibitem{kraus}
Kraus, K. Complementary observables and uncertainty relations. {\it Phys. Rev. D} {\bf 1987}, {\it 35}, 3070--3075.

\bibitem{maass}
Maassen, H.; Uffink, J.B.M. Generalized entropic uncertainty relations. {\it Phys. Rev. Lett.} {\bf 1988}, {\it 60}, 1103--1106.

\bibitem{cancon16}
Rastegin, A.E. Entropic uncertainty relations for successive
measurements of canonically conjugate observables. {\it Ann. Phys.
(Berlin)} {\bf 2016}, {\it 528}, 835--844.

\bibitem{gegen17}
Rastegin, A.E. On entropic uncertainty relations in the presence of a minimal length. {\it Ann. Phys.} {\bf 2017}, {\it 382}, 170--180.

\bibitem{pvb90}
Pegg, D.T., Vaccaro, J.A., Barnett, S.M. Quantum-optical phase and canonical conjugation. {\it J. Mod. Opt.} {\bf 1990}, {\it 37}, 1703--1710.

\bibitem{gvb95}
Gonzalez, A.R.; Vaccaro, J.A.; Barnett, S.M. Entropic uncertainty
relations for canonically conjugate operators. {\it Phys. Lett. A}
{\bf 1995}, {\it 205}, 247--254.

\bibitem{renyi61}
R\'{e}nyi, A. On measures of entropy and information. In {\it
Proceedings of the 4th Berkeley Symposium on Mathematical
Statistics and Probability.} University of California Press,
Berkeley {\bf 1961}, 547--561.

\bibitem{jizba}
Jizba, P.; Arimitsu, T. The world according to R\'{e}nyi: thermodynamics of multifractal systems. {\it Ann. Phys.} {\bf 2004}, {\it 312}, 17--59.

\bibitem{tsallis}
Tsallis, C. Possible generalization of Boltzmann--Gibbs statistics. {\it J. Stat. Phys.} {\bf 1988}, {\it 52}, 479--487.

\bibitem{imrec08}
Csisz\'{a}r, I. Axiomatic characterizations of information measures. {\it Entropy} {\bf 2008}, {\it 10}, 261--273.

\bibitem{bengtsson}
Bengtsson, I.; \.{Z}yczkowski, K. Geometry of Quantum States: An
Introduction to Quantum Entanglement. {\it Cambridge University
Press, Cambridge} {\bf 2006}.

\bibitem{hbb15}
Holik, F.; Bosyk, G.M.; Bellomo, G. Quantum information as a
non-Kolmogorovian generalization of Shannon's theory. {\it
Entropy} {\bf 2015}, {\it 17}, 7349--7373.

\bibitem{bzhpl16}
Bosyk, G.M.; Zozor, S.; Holik, F.; Portesi, M.; Lamberti, P.W. A
family of generalized quantum entropies: definition and
properties. {\it Quantum Inf. Process.} {\bf 2016}, {\it 15},
3393--3420.

\bibitem{IBB84}
Bia{\l}ynicki-Birula, I. Entropic uncertainty relations. {\it Phys. Lett. A} {\bf 1984}, {\it 103}, 253--254.

\bibitem{IBB06}
Bia{\l}ynicki-Birula, I. Formulation of the uncertainty relations in terms of the R\'{e}nyi entropies. {\it Phys. Rev. A} {\bf 2006}, {\it 74}, 052101.

\bibitem{mbg02}
de la Madrid, R.; Bohm, A.; Gadella, M. Rigged Hilbert space treatment of continuous spectrum. {\it Fortschr. Phys.} {\bf 2002}, {\it 50}, 185--216.

\bibitem{blokh73}
Blokhintsev, D.I. Space and Time in the Microworld. {\it D. Reidel Publishing Company, Dordrecht} {\bf 1973}.

\bibitem{huang12}
Huang, Y. Variance-based uncertainty relations. {\it Phys. Rev. A} {\bf 2012}, {\it 86}, 024101.

\bibitem{mpati14}
Maccone, L.; Pati, A.K. Stronger uncertainty relations for all incompatible observables. {\it Phys. Rev. Lett.} {\bf 2014}, {\it 113}, 260401.

\bibitem{prz13}
Pucha{\l}a, Z.; Rudnicki, {\L}.; \.{Z}yczkowski, K. Majorization
entropic uncertainty relations. {\it J. Phys. A: Math. Theor.}
{\bf 2013}, {\it 46}, 272002.

\bibitem{fgg13}
Friedland, S.; Gheorghiu, V.; Gour, G. Universal uncertainty relations. {\it Phys. Rev. Lett.} {\bf 2013}, {\it 111}, 230401.

\bibitem{rpz14}
Rudnicki, {\L}.; Pucha{\l}a, Z.; \.{Z}yczkowski, K. Strong majorization entropic uncertainty relations. {\it Phys. Rev. A} {\bf 2014}, {\it 89}, 052115.

\bibitem{lrud15}
Rudnicki, {\L}. Majorization approach to entropic uncertainty relations for coarse-grained observables. {\it Phys. Rev. A} {\bf 2015}, {\it 91}, 032123.

\bibitem{arkz16}
Rastegin, A.E.; \.{Z}yczkowski, K. Majorization entropic
uncertainty relations for quantum operations. {\it J. Phys. A:
Math. Theor.} {\bf 2016}, {\it 49}, 355301.

\bibitem{ktw14}
Kaniewski, J.; Tomamichel, M.; Wehner, S. Entropic uncertainty from effective anticommutators. {\it Phys. Rev. A} {\bf 2014}, {\it 90}, 012332.

\bibitem{lbp16}
Luis, A.; Bosyk, G.M.; Portesi, M. Entropic measures of joint
uncertainty: effects of lack of majorization. {\it Physica A} {\bf
2016}, {\it 444}, 905--913.

\bibitem{bdqm02}
Basdevant, J.-L.; Dalibard, J. Quantum Mechanics. {\it Springer, Berlin} {\bf 2002}.

\bibitem{bcp14}
Baumgratz, T.; Cramer, M.; Plenio, M.B. Quantifying coherence. {\it Phys. Rev. Lett.} {\bf 2014}, {\it 113}, 140401.

\bibitem{rast16a}
Rastegin, A.E. Quantum-coherence quantifiers based on the Tsallis relative $\alpha$ entropies. {\it Phys. Rev. A} {\bf 2016}, {\it 93}, 032136.

\bibitem{zhc17}
Zhu, H.; Hayashi, M; Chen, L. Coherence and entanglement measures
based on R\'{e}nyi relative entropies. {\it J. Phys. A: Math.
Theor.} {\bf 2017}, {\it 50}, 475303.

\bibitem{luders51}
L\"{u}ders, G. \"{U}ber die zustands\"{a}nderung durch den me{\ss}proze\ss. {\it Ann. Phys. (Leipzig)} {\bf 1950}, {\it 443}, 322--328.

\bibitem{tma12}
Teixeira, A.; Matos, A.; Antunes, L. Conditional R\'{e}nyi entropies. {\it IEEE Trans. Inf. Theory} {\bf 2012},{\it 58}, 4273--4277.

\bibitem{sf06}
Furuichi, S. Information-theoretical properties of Tsallis entropies. {\it J. Math. Phys.} {\bf 2006}, {\it 47}, 023302.

\bibitem{rastkyb}
Rastegin, A.E. Convexity inequalities for estimating generalized conditional entropies from below. {\it Kybernetika} {\bf 2012}, {\it 48}, 242--253.

\bibitem{rastit}
Rastegin, A.E. Further results on generalized conditional entropies. {\it RAIRO--Theor. Inf. Appl.} {\bf 2015}, {\it 49}, 67--92.

\bibitem{BP89}
Barnett, S.M.; Pegg, D.T. On the Hermitian optical phase operator. {\it J. Mod. Opt.} {\bf 1989}, {\it 36}, 7--19.

\bibitem{PB89}
Pegg, D.T.; Barnett, S.M. Phase properties of the quantized single-mode electromagnetic field. {\it Phys. Rev. A} {\bf 1989}, {\it 39}, 1665--1675.

\bibitem{abe92}
Abe, S. Information-entropic uncertainty in the measurements of
photon number and phase in optical states. {\it Phys. Lett. A}
{\bf 1992}, {\it 166}, 163--167.

\bibitem{rastnum12}
Rastegin, A.E. Number-phase uncertainty relations in terms of generalized entropies. {\it Quantum Inf. Comput.} {\bf 2012}, {\it 12}, 0743--0762.

\bibitem{kaw17}
Hsu, L.-Y.; Kawamoto, S.; Wen, W.-Y. Entropic uncertainty relation
based on generalized uncertainty principle. {\it Mod. Phys. Lett.
A} {\bf 2017}, {\it 32}, 1750145.

\end{thebibliography}
\end{document}